\documentstyle[12pt]{article}

\catcode`\@=11
\long\def\@makefntext#1{
\protect\noindent \hbox to 3.2pt {\hskip-.9pt
$^{{\ninerm\@thefnmark}}$\hfil}#1\hfill}                %CAN BE USED

\def\@makefnmark{\hbox to 0pt{$^{\@thefnmark}$\hss}}  %ORIGINAL

\def\ps@myheadings{\let\@mkboth\@gobbletwo
\def\@oddhead{\hbox{}
\rightmark\hfil\ninerm\thepage}
\def\@oddfoot{}\def\@evenhead{\ninerm\thepage\hfil
\leftmark\hbox{}}\def\@evenfoot{}
\def\sectionmark##1{}\def\subsectionmark##1{}}

\renewcommand{\thefootnote}{\fnsymbol{footnote}}

\newcounter{sectionc}\newcounter{subsectionc}\newcounter{subsubsectionc}
\renewcommand{\section}[1] {\vspace*{0.6cm}\addtocounter{sectionc}{1}
\setcounter{subsectionc}{0}\setcounter{subsubsectionc}{0}\noindent
        {\normalsize\bf\thesectionc. #1}\par\vspace*{0.4cm}}
\renewcommand{\subsection}[1] {\vspace*{0.6cm}\addtocounter{subsectionc}{1}
        \setcounter{subsubsectionc}{0}\noindent
        {\normalsize\it\thesectionc.\thesubsectionc. #1}\par\vspace*{0.4cm}}
\renewcommand{\subsubsection}[1]
{\vspace*{0.6cm}\addtocounter{subsubsectionc}{1}
        \noindent
{\normalsize\rm\thesectionc.\thesubsectionc.\thesubsubsectionc.
        #1}\par\vspace*{0.4cm}}

\newcounter{appendixc}
\newcounter{subappendixc}[appendixc]
\newcounter{subsubappendixc}[subappendixc]

\renewcommand{\appendix}[1] {\vspace*{0.6cm}
        \refstepcounter{appendixc}
        \setcounter{figure}{0}
        \setcounter{table}{0}
        \setcounter{equation}{0}
        \renewcommand{\thefigure}{\Alph{appendixc}.\arabic{figure}}
        \renewcommand{\thetable}{\Alph{appendixc}.\arabic{table}}
        \renewcommand{\theappendixc}{\Alph{appendixc}}
        \renewcommand{\theequation}{\Alph{appendixc}.\arabic{equation}}
        \noindent{\bf Appendix \theappendixc #1}\par\vspace*{0.4cm}}

\def\abstracts#1{{

\centering{\begin{minipage}{12.2truecm}\footnotesize\baselineskip=12pt\noindent
        \centerline{\footnotesize ABSTRACT}\vspace*{0.3cm}
        \parindent=0pt #1
        \end{minipage}}\par}}

\renewenvironment{thebibliography}[1]
        {\begin{list}{\arabic{enumi}.}
        {\usecounter{enumi}\setlength{\parsep}{0pt}
\setlength{\leftmargin 1.25cm}{\rightmargin 0pt}
         \setlength{\itemsep}{0pt} \settowidth
        {\labelwidth}{#1.}\sloppy}}{\end{list}}

\newcounter{itemlistc}
\newcounter{romanlistc}
\newcounter{alphlistc}
\newcounter{arabiclistc}

\newcommand{\fcaption}[1]{
        \refstepcounter{figure}
        \setbox\@tempboxa = \hbox{\footnotesize Fig.~\thefigure. #1}
        \ifdim \wd\@tempboxa > 6in
           {\begin{center}
        \parbox{6in}{\footnotesize\baselineskip=12pt Fig.~\thefigure. #1}
            \end{center}}
        \else
             {\begin{center}
             {\footnotesize Fig.~\thefigure. #1}
              \end{center}}
        \fi}

\newcommand{\tcaption}[1]{
        \refstepcounter{table}
        \setbox\@tempboxa = \hbox{\footnotesize Table~\thetable. #1}
        \ifdim \wd\@tempboxa > 6in
           {\begin{center}
        \parbox{6in}{\footnotesize\baselineskip=12pt Table~\thetable. #1}
            \end{center}}
        \else
             {\begin{center}
             {\footnotesize Table~\thetable. #1}
              \end{center}}
        \fi}

\def\@citex[#1]#2{\if@filesw\immediate\write\@auxout
        {\string\citation{#2}}\fi
\def\@citea{}\@cite{\@for\@citeb:=#2\do
        {\@citea\def\@citea{,}\@ifundefined
        {b@\@citeb}{{\bf ?}\@warning
        {Citation `\@citeb' on page \thepage \space undefined}}
        {\csname b@\@citeb\endcsname}}}{#1}}

\newif\if@cghi
\def\cite{\@cghitrue\@ifnextchar [{\@tempswatrue
        \@citex}{\@tempswafalse\@citex[]}}
\def\citelow{\@cghifalse\@ifnextchar [{\@tempswatrue
        \@citex}{\@tempswafalse\@citex[]}}
\def\@cite#1#2{{$\null^{#1}$\if@tempswa\typeout
        {IJCGA warning: optional citation argument
        ignored: `#2'} \fi}}

 1
 1
 1

\font\ninerm=cmr9

\begin{document}
\begin{flushright}
TKU-95-1 July 1995
\end{flushright}
\vskip 12pt
\centerline{\normalsize\bf SPECTROSCOPY AND DECAYS OF HEAVY MESONS
\footnote{A talk given by T.M. at WEIN'95 Conference held at RCNP of
Osaka University, June 1995}}
\baselineskip=22pt

%\vfill
%\vspace*{0.6cm}
\centerline{\footnotesize TAKAYUKI MATSUKI}
\baselineskip=13pt
\centerline{\footnotesize\it  Tokyo Kasei University, 1-18-1 Kaga, Itabashi,
Tokyo 173, JAPAN}
\baselineskip=12pt
\centerline{\footnotesize E-mail: matsuki@ins.u-tokyo.ac.jp}
\vspace*{0.3cm}
\centerline{\footnotesize and}
\vspace*{0.3cm}
\centerline{\footnotesize TOSHIYUKI MORII}
\baselineskip=13pt
\centerline{\footnotesize\it Faculty of Human Development, Kobe University,
 3-11 Tsurukabuto, Nada}
\baselineskip=12pt
\centerline{\footnotesize\it Kobe 657, JAPAN}
\centerline{\footnotesize E-mail: morii@kobe-u.ac.jp}

%\vfill
\vspace*{0.3cm}
\abstracts{
Assuming Coulomb-like as well as confining scalar potential, we have solved
Shr\"odinger equation perturbatively in $1/m_Q$ with a heavy quark mass $m_Q$.
The lowest order equation is examined carefully. Mass levels are fitted with
experimental data for $D/B$ mesons at each level of perturbation. Meson wave
functions obtained thereby can be used to calculate ordinary form factors
as well as Isgur-Wise functions for semileptonic weak decays and other
physical quantities.
All the above calculations
are expanded in $1/m_Q$ order by order to determine parameters as well as to
compare with results of Heavy Quark Effective Theory.}

%\vspace*{0.6cm}
\normalsize\baselineskip=15pt
\setcounter{footnote}{0}
\renewcommand{\thefootnote}{\alph{footnote}}
\section{Introduction}
HQET (Heavy Quark Effective Theory),$^1$ is applied to
many aspects of high energy theories and many kinds of physical
quantities of QCD which can be perturbatively calculated in $1/m_Q$.
Especially those regarding to $B$ meson physics, e.g.,
the Isgur-Wise function of
semileptonic weak deacy processes $B \rightarrow D \ell \nu$ and
the Kobayashi-Maskawa matrix element $V_{cb}$, have been calculated
by many people.
However, applications of HQET to higher order perturbative calculations
are limited only to obtain forms of
higher order operators, and their coefficients should be obtained
so that results be fitted with experimental data.
This is because most of the calculations based on HQET do not introduce
heavy meson wave functions and hence
there is no way to determine those coefficients within the model.

In this paper, we would like to start from introducing
phenomenological dynamics, i.e.,
assuming Coulomb-like vector and confining scalar potential to
$Q \bar q$ bound states (heavy mesons), expand a hamiltonian
in $1/m_Q$ then perturbatively solve Shr\"odinger equation
in $1/m_Q$. Angular part
of the lowest order wave function is exactly solved.
After extracting asymptotic forms of the lowest order wave function
at both $r \rightarrow 0$ and $r \rightarrow \infty$
and adopting the variational method, we numerically obtain
radial part of the trial wave function
which is expanded in polynomials of radial variable $r$.
Then fitting the smallest eigenvalues
of a hamiltonian with masses of $D$ and $D^*$ mesons, a strong
coupling $\alpha_s$ and other potential parameters are determined uniquely.
Using parameters obtained this way, other mass levels are calculated and
compared with experimental data for $D/B$ mesons at each level
of perturbation. Meson wave functions obtained thereby
may be used to calculate ordinary form factors/Isgur-Wise
functions for semileptonic weak decay processes and
other physical quantities as well.

\section{Hamiltonian and Eigenvalue Equation}
Our hamiltonian is given by~\cite{Kobe}
\begin{eqnarray*}
H&=&\left( {\vec \alpha _q\cdot \vec p_q
+\beta _qm_q} \right) +\left(
{\vec \alpha _Q\cdot \vec p_Q
+\beta _Qm_Q} \right) +\beta _q\beta _QS(r) \\
&+& \left\{ {1-{1 \over 2}\left[ {\vec \alpha _q\cdot
\vec \alpha _Q+\left( {\vec \alpha _q\cdot \vec n} \right)\left( {\vec \alpha
_Q\cdot \vec n} \right)} \right]} \right\}V(r).
\end{eqnarray*}
When a heavy quark, $Q$, is treated as a non-relativistic
particle, the hamiltonian is reduced into a $4 \times 2$
matrix operator and the Schr\"odinger equation becomes
\[\left( {H_{-1}+H_0+H_1+H_2+\cdots} \right)\phi =M\phi=(m_Q+E_b)\phi\]
where $m_q-E_b$ is a binding energy and scalar and vector
potentials are given by
\[S\left( r \right)={r \over {a^2}}+b, \quad V\left( r \right)
=-{4 \over 3}{\alpha _s \over r},~{\rm and}\quad
E_b=E_0+\left(m_q \over m_Q\right)~E_1+\cdots.\]
and
\begin{eqnarray*}
  H_{-1}& =&m_Q\\
  H_0& =&\vec \alpha _q\cdot \vec
p+\beta _q\left( {m_q+S} \right)+V.
\end{eqnarray*}
Higher order terms like $H_1$ and $H_2$ have some complicated stuructures
and are not described here for simplicity. Subscripts $i$ of
$H_i$ denote the order of $1/m_Q$.

The -1st order eigenvalue equation is given by
$H_{-1}\phi_0=m_Q \phi_0,$ which is a trivial one.
The next 0-th order becomes a non-trivial equation,
$ H_0  \Psi_{j\,m}^k = E_k^0 \Psi_{j\,m}^k$ ,
which can be reduced into
\[\left( {\matrix{{m_q+S+V}&{-\partial _r+{k \over r}}\cr
{\partial _r+{k \over r}}&{-m_q-S+V}\cr
}} \right)\left( {\matrix{{u_k\left( r \right)}\cr
{v_k\left( r \right)}\cr
}} \right)=E_k^0\left( {\matrix{{u_k\left( r \right)}\cr
{v_k\left( r \right)}\cr
}} \right).\]
This equation is numerically solved by taking into
account the asymptotic behaviors at both $r\to0$ and $r\to\infty$
and their forms are given by
\[u_k(r), v_k(r) \sim w(r)\; r^{\gamma} \exp\left(-(m_q+b) r -{1\over 2}
\left(r\over a\right)^2\right),\]
where $\gamma=\sqrt{k^2-(4\;\alpha_s/3)^2}$ and $w(r)$ is some polynomial of
$r$.
In the above
an eigenfunction, $\Psi_{j\,m}^k$, can, in general, be given by
\[\Psi _{j\,m}^k(\vec r,\Omega)=\Psi _k(r)\;y_{j\,m}^k(\Omega)=
{1\over r}\;\left( {\matrix{u_k( r)\cr -i v_k( r)
\left(\vec\sigma_q\cdot\vec n\right)\cr
}} \right)\;y_{j\,m}^k (\Omega),\]
where the function $y_{j\,m}^k (\Omega)$ is a linear combination of
$Y_{j m}$ and vector spherical harmonics
$\vec\sigma\cdot\vec X_{j j}^m(\Omega)$
and $\vec\sigma \cdot \vec Y_{j\,(\pm)}^m(\Omega)$,
which satisfies
\[(\vec\sigma_q \cdot \vec\ell)\;
y_{j\;m}^k(\Omega)=-(k+1)\;y_{j\;m}^k(\Omega)\]
with $k=\pm j,\;{\rm or}\; \pm (j+1)$ and $\vec\ell$ is an angular
momentum. The first few series of $y_{j m}^k$ functions are given by
\[y_{0\;0}^{1}=-{1\over \sqrt{4\pi}}\left(\vec n\cdot\vec\sigma\right),\quad
  y_{0\;0}^{-1}={1\over \sqrt{4\pi}},\]
\[y_{1 0}^1={-i\over \sqrt{4\pi}}(\vec n\cdot \vec\sigma)\,\sigma_z,\quad
y_{1 1}^1={-i\over \sqrt{2\pi}}(\vec n\cdot \vec\sigma)\,\sigma_+,\quad
y_{1 -1}^1={-i\over \sqrt{2\pi}}(\vec n\cdot \vec\sigma)\,\sigma_-, \]
where $y_{j m}^{-1}=-(\vec n\cdot \vec\sigma)\,y_{j m}^1$,
$\sigma_\pm=(\sigma_x\pm i\sigma_y)/2$, and $\vec n=\vec r/r$.

Matrix elements of interaction terms among eigenfunctions
are expressed as below. Degeneracy can be resolved
by diagonal elements of $H_1$ and $H_2$ with respect to
the $k$ quantum number. Inclusion of off-diagonal elements of $H_1$ and $H_2$
are absorbed into wave function corrections.
Calculating all the matrix elements from the hamiltonian given above,
total mass matrix is given by,
\[\left( {\matrix{{\scriptstyle E_{-1}+E_{-1,0}}
&0&0&0&0&0&0&0\cr
0&{\scriptstyle E_{-1}+E_{-1,1}}
&0&0&0&0&{\scriptstyle V_{-1,2}}&0\cr
0&0&{\scriptstyle E_1+E_{1,0}}&0&0&0&0&0\cr
0&0&0&{\scriptstyle E_1+E_{1,1}}&{\scriptstyle V_{1,-2}}&0&0&0\cr
0&0&0&{\scriptstyle V_{-2,1}}&{\scriptstyle E_{-2}+E_{-2,1}}&0&0&0\cr
0&0&0&0&0&{\scriptstyle E_{-2}+E_{-2,2}}&0&0\cr
0&{\scriptstyle V_{2,-1}}&0&0&0&0&{\scriptstyle E_2+E_{2,1}}&0\cr
0&0&0&0&0&0&0&{\scriptstyle E_2+E_{2,2}}\cr
}} \right),\]
where
\[E_k=m_Q+E_k^{(0)}+E_k^{(1)}+E_k^{(2)},
\quad E_{k,\;j}=E_{k,\;j}^{(1)}+E_{k,\;j}^{(2)},\quad
  V_{k,\;k'}=V_{k,\;k'}^{(1)}+V_{k,\;k'}^{(2)},\]
superscripts mean the order of $1/m_Q$, $k$ and $k'$ stand for
$k$ quantum number, and $j$ for the total angular momentum.
%For instance, $V_{k,\;k'}^{(2)}$ means the matrix element of some terms
%of $H_2$ between $\Psi_{j\;m}^k$ and $\Psi_{j\;m}^{k'}$.

Corrections to wavefunctions and eigenvalues can be calculated
by applying the ordinary perturbation.
For instance, in the above $8\times 8$ mass matrix up to $O(1/m_c)$,
the wave function for $(2,2)$ element is given by
\begin{eqnarray*}
\Psi_2&=&\Psi^{-1}_{1 m}(r,\Omega)-7.92\times~
  10^{-3} \Psi^2_{1 m} (r, \Omega)\\
  &=& \Psi^{-1}_{1 m}(r,\Omega)-1.19\times {\left( m_u \over m_c \right)}
  \Psi^2_{1 m} (r, \Omega),
\end{eqnarray*}
where
\[\alpha_s={g_s^2\over 4\pi}=0.299,\quad a=2.10~{\rm GeV}^{-1},
\quad b=-0.084~{\rm GeV},\]
\[m_u=0.01 ~{\rm GeV},\quad m_c = 1.461 ~{\rm GeV},
\quad m_b = 4.92 ~{\rm GeV}.\]
If $D$ and $D^*$ are identified as $\Psi_{00}^{-1}$ and $\Psi_{1m}^{-1}$'
respectively, $D$ and $B$ meson masses are calculated to be
\[m_D=1.867\;{\rm GeV},\quad m_{D^*}=1.9241\;{\rm GeV},\quad
  m_B=5.279\;{\rm GeV},\quad m_{B^*}=5.298\;{\rm GeV}.\]

\section{ Comments and Discussions}
One can easily see degeneracy among the lowest lying pseudoscalar
and vector states as follows. Define
\[\left| P \right\rangle =\left| D^\pm ,\;D^0\right>=\Psi _{0\;0}^{-1},
  \quad \left| {V,\;m } \right\rangle =\left| D^*\right>=
\Psi _{1\;m}^{-1},\]
where $\Psi _{j\;m}^k$ is an eigenfunction obtained in the last chapter
and quantum number $k$
can take only $\pm j$, or $\pm \left( {j+1} \right)$.
Since these states have the same quantum number $k=-1$,
these have the same masses as well as the same wave functions
up to the 0-th order calculation in $1/m_Q$.
One needs to develop perturbation of energy and wave function for each state
in terms of $1/m_Q$ to obtain higher order corrections.

Next we would like to discuss qualitative features of form factors
functions. Let us think about to calculate form factors
for semileptonic decay of $B \rightarrow D \ell \nu$.
Taking a simple form for the lowest lying wave function both for $B$ and $D$ as
\[\Psi^{1S} \sim e^{-b^2\;r^2/2},\]
where a parameter $b$ is determined by a variational principle,
$\delta(\Psi^{1S\;\dagger}H\Psi^{1S})=0$.
Then form factors/Isgure-Wise functions are
given by
\[ F(q^2) \sim \exp\left[ {\rm const.}\;m_q^2\; (q^2-q^2_{\max})\right],
\;{\rm or}\quad
\xi(\omega) \sim \exp\left[ {\rm const.}\;m_q^2\; (1-\omega)\right],\]
where
\[q^2=(p_B-p_D)^2,\quad \omega=v_B \cdot v_D,\quad
q^2_{\max}=(m_B-m_D)^2 \leftrightarrow \omega_{\max}=1.\]
This means behavior of form factors strongly depends on a value of light
quark mass $m_q$. One needs current quark mass for $m_q$ to
reproduce heavy meson mass spectrum, while constituent quark mass is apparently
used to calculate form factors in all published papers.
In order to explain this situation, one is forced to use running
mass~\cite{Pol} $m_q(q^2)$ for light quark mass.

\end{document}